\documentclass{an}
\usepackage{graphicx}
\usepackage{times}
\usepackage{fancyhdr}
\newcommand{\lsim}{\,\lower2truept\hbox{${<\atop\hbox{\raise4truept\hbox{$\sim$}}}$}\,}
\newcommand{\gsim}{\,\lower2truept\hbox{${>\atop\hbox{\raise4truept\hbox{$\sim$}}}$}\,}

\def\Ohat{\hat{\Omega}_b}

\def\tsu27{\left(\frac{T_0}{2.7K}\right)}

\begin{document}

\title{Cyclotron emission effect on CMB spectral distortions}

\author{C.~Burigana and A.~Zizzo}
\institute{INAF-IASF Bologna, Via Gobetti 101, 40129 
Bologna, Italy}

\date{Received; accepted; published online}

\abstract{We investigated the role of the cyclotron emission (CE)
associated to cosmic magnetic fields (MF) on the evolution of
cosmic microwave background (CMB) spectral
distortions. We computed the photon and energy injection rates by
including spontaneous and stimulated emission and absorption. These
CE rates have been compared with those of
bremsstrahlung (BR) and double
Compton scattering (DC), for realistic CMB distorted spectra at various
cosmic epochs. For reasonable MF strengths we found that the
CE contribution to the evolution of the CMB spectrum is much smaller than 
the BR and DC contributions.
The constraints on
the energy exchanges at various redshifts can be then derived, under 
quite
general assumptions, by considering only Compton scattering (CS),
BR, and DC, other than
the considered dissipation process.
Upper limits to the CMB polarization degree induced by CE
have been estimated.
}

\correspondence{burigana@bo.iasf.cnr.it}

\maketitle

\section{Introduction}
\label{sec:intro}

The 
%cosmic microwave background (CMB) 
CMB 
spectrum emerged from the 
thermalization redshift,
$z_{therm}$,
with a shape very close to a 
blackbody (BB) 
one, owing to the tight coupling between radiation and matter through
%Compton scattering (CS) 
CS
and photon production/absorption processes,
%double Compton scattering (DC) and bremsstrahlung (BR). 
DC and BR.
In the presence of a cosmic MF, another  
photon production/absorption process,
%the cyclotron emission (CE),
CE,
operates in the cosmic plasma
(Puy \& Peter 1998).
Since the very large electron conductivity
the magnetic flux through any 
loop moving with fluid is a 
conserved quantity.
On scale where diffusion can be neglected the field is said to be %
{\it frozen-in}, in the sense that lines of force move together with %
the fluid. Assuming that the Universe expands isotropically, 
magnetic flux conservation implies 
%
%\begin{equation}\label{eq:evoluzionediB}
${\bf B}(t)={\bf B}(t_0)\left(\frac{a(t_0)}{a(t)}\right)^2={\bf 
B}_0(1+z)^2 \, $,
%\end{equation}
%
${\bf B}_0$ being the MF at the present time.
Considering the cyclotron spontaneous emission,
absorption and stimulated emission terms (Afshordi 2002),
we derived the CE contribution 
to the evolution of the CMB photon occupation number, $\eta$, 
as a further term 
in the Kompaneets equation
(Zizzo \& Burigana 2005):

\vskip -0.4cm
\begin{eqnarray}\label{eq:kompaneets+ciclotrone}
\frac{\partial\eta}{\partial t}&=&\frac{1}{\phi}\frac{1}{t_C}
\frac{1}{x^2}\frac{\partial}{\partial x}\left[x^4\left[\phi
\frac{\partial\eta}{\partial x}+\eta(1+\eta)\right]\right] \nonumber\\
&+&\left[K_{BR}\frac{g_{BR}}{x_e^3}e^{-x_e}+K_{DC}\frac{g_{DC}}{x_e^3}+K_{CE}
\delta(x_e-x_{e,CE})\right] \nonumber\\
&\cdot&\left[1-\eta(e^{x_e}-1)\right] \, ; \nonumber
\end{eqnarray}
\vskip -0.3cm

\noindent
here the BR and DC rates are defined by the coefficients $K(z)$ and the 
Gaunt factors, 
$g_{BR}$ and $g_{DC}$, 
%\cite{karzaslatter,rybickilightman}
%\cite{gould84}
%are given in 
%Burigana et al. (1991) and Burigana et al. (1995),
%\cite{buriganaetal91a,buriganaetal95},
$t_C=mc^2/[kT_e(n_e \sigma_T c)]$ 
is the timescale for the achievement of kinetic equilibrium between
radiation and matter,
$\phi=T_e/T_r$ 
where $T_r=T_0(1+z)$ is the CMB temperature and $T_e$ the electron one, 
$x=\phi x_e$ (see, e.g., Burigana, De Zotti \& Danese 1995 and 
references therein), 
$x_{e,CE}={h\nu_c \over kT_e} \simeq 
4.97\times 10^{-5}\cdot \phi^{-1}\tsu27 ^{-1}B_0(1+z) \, $
is the dimensionless frequency of CE,
and 
$K_{CE}(z)=\frac{4\pi^2\,e\,c}{3 B(z)}\,n_e=4.64\times10^{-4}\,
\Ohat\cdot B^{-1}(z)(1+z)^3\mbox{ s}^{-1} \, $ where
$\Ohat =\Omega_b [H_{0}/(50 {\rm Km} {\rm s}^{-1} {\rm Mpc}^{-1})]^2 \,$,
$\Omega_b$ being 
the baryon density parameter and $H_0$ the Hubble constant.

In Fig.~1 we compare $x_{e,CE}$ with the 
frequency $x_{e,c}$,
at which the combined DC and BR absorption time
$t_{abs}$ is equal to $t_C$ and 
the maximum frequency $x_{e,abs}$, 
at which DC and BR could re-establish a BB spectrum after a 
distortion.

\begin{figure}
\vskip -0.3cm
\resizebox{\hsize}{!}
{\includegraphics[]{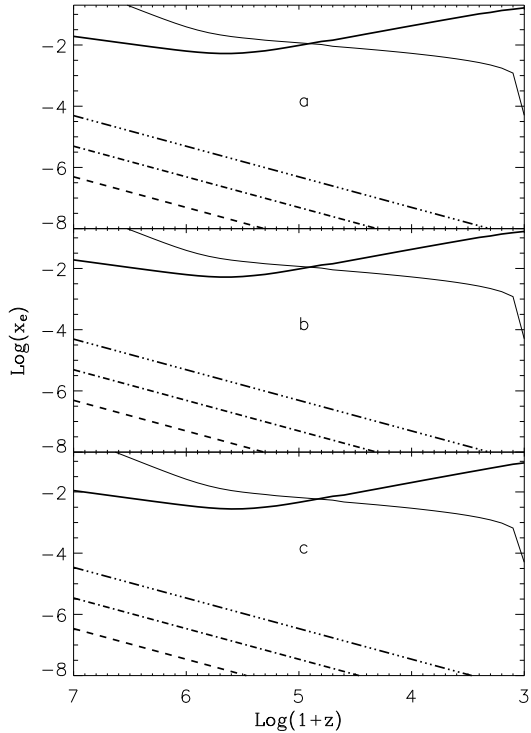}}
\vskip -0.3cm
\caption{Comparison between the characteristic frequencies $x_{e,c}$
(thick solid line) $x_{e,abs}$ (thin solid line), and $x_{e,CE}$ for
different values of the MF $B_0$:
$10^{-9}$G (thick dashed line), $10^{-8}$G
(thick dot-dashed line), $10^{-7}$G (thick three dots and dash line).
Different values of the fractional energy
$\Delta\varepsilon/\varepsilon_i$
($\simeq \mu_0/1.4$ for
$\mu_0 \ll 1$),
injected in the CMB radiation field 
in the case of early (BE-like) distorted spectra,
have been assumed in the panels:
$10^{-4}$
(panel {\it a}), $10^{-2}$ (panel {\it b}), $1$ (panel {\it c}).
The cosmological parameters have been assumed in agreement with WMAP (see 
Tab.~3
of Bennett et al. 2003);
%\cite{bennett});
nevertheless 
%the dependence on 
the detailed choice of them
is not 
%particularly 
critical here.}
\label{fig1}
\end{figure}

\section{Results}
\label{sec:results}

We applied the above formalism for realistic assumptions of $\eta$
to compute the global photon production rate as a function of the relevant
parameters and understand the
role of the CE in the thermalization 
and evolution of the CMB spectrum.
We computed the photon number 
%production rate
%$$\left(\frac{d n}{d t}\right)=
%8\pi\left(\frac{k_BT_e}{h c}\right)^3\int 
%\left(\frac{\partial\eta(x_e)}%
%{\partial t}\right)x_e^2dx_e $$
and photon energy production rates. 
%$$\left(\frac{d \varepsilon}{d t}\right)=
%\frac{8\pi}{h^3 c^3}(k_BT_e)^4\int 
%\left(\frac{\partial\eta(x_e)}%
%{\partial t}\right)x_e^3dx_e $$.
The contributions by BR, DC, and CE 
for an energy dissipation occurring at high $z$ 
by assuming %
for $\eta$ 
%(i.e. in
%Eqs. (\ref{eq:productionrates_cy}) and (\ref{eq:ratesdibredc}))
a pure Bose-Einstein (BE) formula with a constant chemical potential
($\mu=\mu_0$) or a 
BE-like spectrum with a frequency dependent chemical potential
($\mu=\mu(x_e)$),
as appropriate during the kinetic equilibrium epoch ($z \gsim z_1$),
are compared in Fig.~2.
Clearly, considering the correct BE-like spectrum
we found that the CE contribution 
to the thermalization process of CMB spectrum
after an early heating is negligible, because of 
the very low frequency location of CE photons 
and the high efficiency of DC and BR to keep the long
wavelength region
of the CMB spectrum close to a BB
during the formation of the spectral distortion. 
We performed a similar analysis in the case of 
Comptonization like distortions, i.e. Comptonization spectra 
corrected at low frequencies 
for the effect of DC and BR as appropriate to represent 
spectra distorted at relatively late epochs, and found 
again a negligible role of the CE term because of 
its very low frequency location.

\begin{figure}
\vskip -1.35cm
\resizebox{\hsize}{!}
{\includegraphics[]{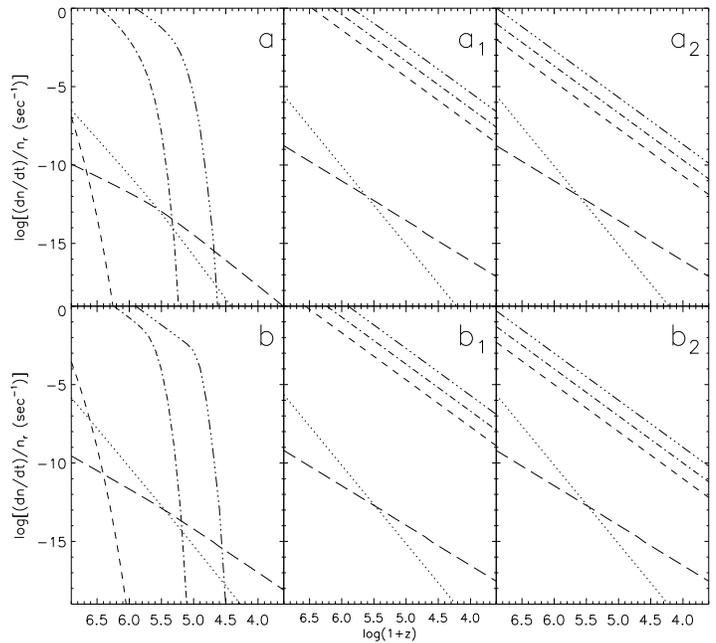}}
\vskip -0.3cm
\caption{Photon number injection 
fractional (i.e. divided by the photon number 
density of a BB at temperature $T_0(1+z)$)
rates at the redshift $z$
from DC (dots), BR (long dashes)
and CE
(dashes: $B_0=2\times10^{-6}$~G;
dots-dashes: $B_0=2\times10^{-5}$~G;
three dots-dashes: $B_0=2\times10^{-4}$~G) in the presence of
an early distortion with $\mu_0= 1.4 \times 10^{-2}$ (top panels)
or $1.4$ (bottom panels)
occuring exactly at the redshift $z$ (properly speaking this result
holds for $z \gsim z_1$ while it is only indicative for
$z \lsim z_1$). In panels $a$ and $b$ 
the appropriate BE-like spectrum ($\mu = \mu (x_e)$) is assumed.
In panels $a_1$ and $b_1$ the simplistic case $\mu = \mu_0$ is considered.
Panels $a_2$ and $b_2$ are identical to panels
$a_1$ and $b_1$ but for
$B_0=10^{-9}$~G (dashes),
$B_0=10^{-8}$~G (dots-dashes),
$B_0=10^{-7}$~G (three dots-dashes): the simplistic 
assumption $\mu = \mu_0$ would imply the (wrong) conclusion of a dominance
of the CE contribution for reasonable values of the MF.
(Note that the DC and BR rates in panels $a_1$, $b_1$, $a_2$, and $b_2$
are only indicative because of the integral low frequency divergency 
in the case of a frequency independent chemical potential.
We corrected here an erroneous multiplicative factor 2 
included in all the rates plotted 
in Figs.~2 and 3 of Zizzo \& Burigana (2005)
because of a typo only in the corresponding IDL 
visualization procedure).}
\label{fig2}
\end{figure}

We derived also upper limits 
to the CMB polarization degree
induced by the CE process. The corresponding signal
turns to be much more smaller than the signal of CMB polarization 
anisotropies and of that 
predicted for the CMB polarization
anisotropies ``directly'' induced by cosmic MFs 
(Subramanian et al. 2003),
and is below observational chances.

For many models of dissipation mechanisms, as in the case of 
dissipation processes mediated 
by energy exchanges between matter and radiation 
or associated to photon injections at $\nu$ significantly
different from $\nu_c$, 
the role of CE 
in the evolution of CMB spectral distortions
is negligible for cosmic MF realistic values.
In particular, it cannot re-establish a BB
spectrum after the generation of early distortions
and it cannot significantly distort the CMB spectrum.
No significant limits on the cosmic MF
strength can be then set by constraints on CMB spectral distortions 
when interpreted as produced by CE.
Consequently, 
energy dissipation processes can be studied 
%, under quite general assumptions,
by considering only CS, BR, and DC, other 
than, obvioulsy, the considered mechanism(s).

\end{document}